\documentclass{ws-rv961x669}
\usepackage{subfigure}   
\usepackage{ws-rv-thm}   
\usepackage{ws-rv-van}   

\usepackage{graphicx}
\usepackage{subfigure}
\usepackage{textcomp}
\usepackage{slashed}
\usepackage{wrapfig}

\def\beq{\begin{equation}}
\def\eeq{\end{equation}}
\def\theta{\vartheta}

\newcommand{\ba}{\begin{eqnarray}}
\newcommand{\ea}{\end{eqnarray}}
\newcommand{\lsim}   {\mathrel{\mathop{\kern 0pt \rlap
  {\raise.2ex\hbox{$<$}}}
  \lower.9ex\hbox{\kern-.190em $\sim$}}}
\newcommand{\gsim}   {\mathrel{\mathop{\kern 0pt \rlap
  {\raise.2ex\hbox{$>$}}}
  \lower.9ex\hbox{\kern-.190em $\sim$}}}

\begin{document}



\title{Neutrino Physics and Astrophysics}

\chapter[Overview]{Neutrino Physics and Astrophysics Overview}

\author[Floyd W. Stecker]{Floyd W. Stecker\footnote{Floyd.W.Stecker@nasa.gov}}
\address{Astrophysics Science Division, NASA Goddard Space Flight Center \\ Greenbelt, MD 20771 USA \\
and Dept. of Physics and Astronomy \\
University of California Los Angeles, Los Angeles, CA 90095 USA\\}

\vspace{0.8cm}
 
     {\it I have done something very bad today by proposing a particle that cannot be detected; it is something no theorist should ever do.}
      
   - Wolfgang Pauli
\vspace{0.2cm}

~
\section{History and Mystery}
~

Despite Pauli's pessimistic pronouncement, his spectral particle eventually was detected, although it took 26 years to do so.
In fact, the {\index neutrino} is the second most abundant particle in the universe. There are approximately three hundred million neutrinos in every cubic meter, relics of the {\index big bang}. The most abundant particle, the photon, is well understood. We have a complete theory of the photon, namely {\index quantum electrodynamics} (QED), developed in the middle of the 20th century by {\index Julian Schwinger}, {\index Richard Feynman}, {\index Freeman Dyson}, and {\index Sin-Itiro Tomonaga}~\cite{Schweber:1994qa}. By contrast, the character of the neutrino is not completely understood to this day. This volume is devoted to the various aspects of the neutrino, both understood and not understood, and the role of the neutrino in our understanding of the Universe. 

The neutrino was first postulated by {\index Wolfgang Pauli} in 1930 in order to explain the puzzling continuous shape of the electron spectrum in {\index $\beta$-decay} observed by {\index James Chadwick} in 1914~\cite{Chadwick:1914zz}. In a letter on December 4th, 1930, he wrote~\cite{Pauli:2000ak}, 

{\it "The continuous beta spectrum would make sense under the assumption that in beta decay, in addition to the electron, a neutron is emitted such that the sum of the energies of neutron and electron is constant."} 

\newpage

If $\beta$-decay were a two-body decay, the electron spectrum would have had a fixed energy. The continuous {\index $\beta$-decay} spectrum strongly suggested to Pauli that a ghostly, electrically neutral, particle had to be involved in the decay in order to conserve energy. Energy conservation is critical to all of physics. According to a powerful theorem of the brilliant mathematician {\index Emmy Noether}, if energy is not conserved, then physics itself would not be independent of time~\cite{Noether:1918zz}. Pauli further pointed out that the neutral particle involved in the decay would have spin-1/2 and obey his {\index exclusion principle}\footnote
{Leptons such as the neutrino and quarks obey {\index  ~Fermi-Dirac statistics} and the {\index 
~ Pauli exclusion principle} and are called {\it fermions}. The gauge particles such as the photon that carry the forces obey {\index Bose-Einstein statistics} and are called {\it bosons}.} The particle that Pauli called a {\it neutron} was renamed by {\index Enrico Fermi} in the following year. Fermi renamed the particle the {\it neutrino}, giving it a well-deserved diminutive Italian designation. 

~
\begin{wrapfigure}{r}{0.5\linewidth}
\begin{center}
  \includegraphics[width=3.5cm]{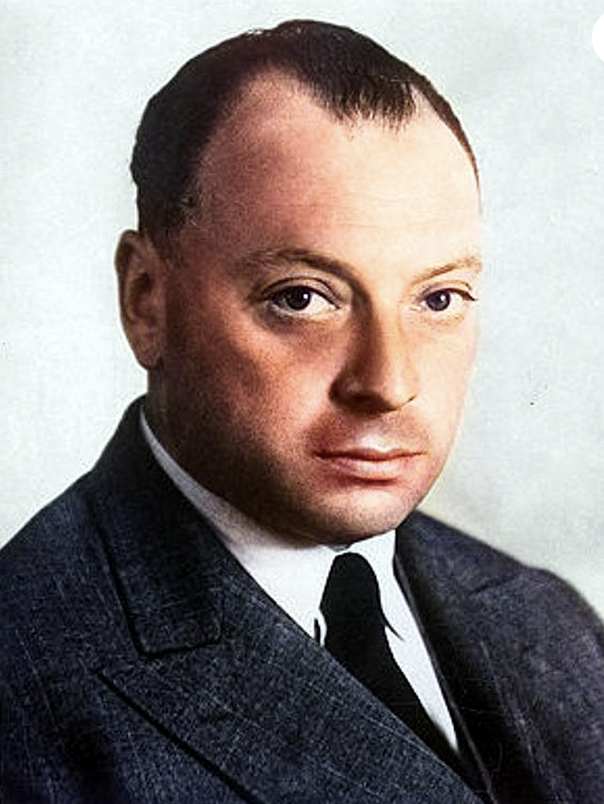}
\end{center}
\centering
 Wolfgang Pauli, the father of the neutrino. 
\end{wrapfigure}

One might expect that neutrinos, being the lightest of the known fundamental particles (except for photons), should be the simplest. Paradoxically, in many ways they appear to be the most complex. Complicating matters is the extremely small cross section for neutrinos interacting with matter. This leads to challenges for their detection. They are so hard to find and that the {\index  electron neutrino} was only discovered in 1956 by {\index Fred Reines} and {\index Clyde Cowan}~\cite{Cowan:1956rrn}.
~
The extremely small cross section of the neutrino interaction with nuclei, and the astronomically long lifetime (by physics standards) of free neutron $\beta$-decay, $\cal{O}$(10$^3$s), are characteristics defining these to be {\it weak} interactions, {\it vis-\`{a}-vis} the {\it strong} hadronic interactions. The epic triumph of weak interaction theory was the unification of weak and electromagnetic interactions~\cite{Aitchison:1982kj} for which {\index Sheldon Glashow}, {\index Stephen Weinberg} and {\index Abdus Salam} received the {\index Nobel Prize} in 1979. This theory, which replaced Enrico Fermi's 1934 theory of {\index weak interactions}~\cite{Fermi:1934hr}, involved the so-called {\index Higgs mechanism} of {\index spontaneous symmetry breaking}, $ SU(2)_L \otimes U(1)_Y \ \rightarrow \ U(1)_{\mathrm{QED}} $, whereby the electroweak symmetry group is broken by the vacuum to the electromagnetic subgroup, defining the masses and roles of the massless photon and three newly created massive {\index gauge bosons}, $W^{+}, W^{-}$ and $Z^{0}$ that mediate the {\index charged current} and {\index neutral current} weak interactions. 

The $W^{+}, W^{-}$ and $Z^{0}$ bosons have masses $\cal{O}$(100) proton masses. The fact that neutrinos only interact weakly via the weak force is related to the heavy masses of the $W^{+}, W^{-}$ and $Z^{0}$ gauge bosons. For example, the decay rate of the $\mu$, is proportional to $M_W^{-4}$. 

Because the weak gauge bosons are so heavy, they were finally only discovered in 1983 with the advent of the Super Proton Synchrotron~\cite{Rubbia:1984xy}. For this discovery, {\index Carlo Rubbia} and {\index Simon van der Meer} were  awarded the Nobel Prize 1984. The final missing key to electroweak unification, the slightly heavier Higgs boson, was discovered in 2012. {\index Fran\c{c}ois Englert} and {\index Peter Higgs} were awarded the {\index Nobel Prize} in 2013 for the Higgs mechanism~\cite{Higgs:1964pj,Englert:1964et}. In all, many theorists contributed to the final development of electroweak theory.  

Since then, the plot of the neutrino story has become more and more complex. Following the discovery of the electron neutrino ($\nu_e$), by Reines and Cowan, two more neutrino {\it flavors} were discovered, the {\index muon neutrino} ($\nu_{\mu}$), discovered in 1962 by {\index Leon Lederman}, {\index Melvin Schwartz} and {\index Jack Steinberger}~\cite{Danby:1962nd}, and the {\index tau neutrino} ($\nu_{\tau}$), discovered by the {\index DONUT} ({\it D}irect {\it O}bservation of the {\it NU}-{\it T}au) collaboration at {\index Fermilab} in 2000~\cite{DONUT:2000fbd}. The three neutrino flavors are paired with the three flavors of charged leptons, viz., the electron (e), the muon ($\mu$), and the tau ($\tau$). The three neutral leptons, viz., the neutrinos, are defined by their weak interactions. They only interact specifically with their charged counterparts. These are referred to as the three lepton families. They are matched by the three {\index quark} families. The reason for three families is presently unknown. In fact, the lightest family of quarks and leptons is the only one that appears to be necessary to describe the everyday universe that we know and love.

Another neutrino puzzle arose when neutrinos from nuclear reactions in the solar core were detected at the {\index Homestake mine} in South Dakota in 1964 by {\index Raymond Davis Jr}. and {\index John Bahcall}~\cite{Bahcall:1964gx}. Their flux was less than half that given by the theoretical predictions. This was known as the {\it solar neutrino problem}. 

The solution to the {\index solar neutrino problem} and similar phenomena involving atmospheric neutrinos lay in the phenomenon that neutrino flavors can mutate into each other. 
The possibility of such oscillations between neutrino flavors exhibiting different weak interactions was postulated by {\index Bruno Pontecorvo} in 1957~\cite{Pontecorvo:1959sn}. This prediction was finally experimentally confirmed by the the {\index Super-Kamiokande Collaboration} in 1998 and the {\index Sudbury Neutrino Observatory Collaboration} in 2001~\cite{Super-Kamiokande:2001ljr,SNO:2002tuh}. Those results showed that the three neutrino flavors do indeed exhibit the quantum phenomenon of {\index neutrino oscillations}. For their discoveries, {\index Takaaki Kajita} and {\index Arthur McDonald} shared the {\index Nobel Prize} in 2015. 

Thus, neutrinos that are produced in one {\index flavor eigenstate} can be detected to be in a different flavor eigenstate after traveling over some finite distance. This is because their {\index mass eigenstate}s are different from their flavor eigenstates. Such oscillations can be described by a 3 x 3 unitary transformation matrix known as the Pontecorvo-Maki-Nakagawa-Sakata ({\index PMNS}) matrix, which is analogous with the Cabibbo-Kobayashi-Maskawa ({\index CKM}) matrix describing the mixing between quark flavors.

The scenario of these ghostly particles transforming into each other is quite mysterious. The mystery is compounded by the fact that the the parameters that define the CKM and PMNS matrices, both of which allow for the violation of {\index $\cal{CP}$} (charge conjugation times parity) symmetry,\footnote{It was shown by Andrei Sakharov in 1991 that violation of $\cal{CP}$ symmetry is a requirement for producing an excess of matter over antimatter in the early universe~\cite{Sakharov:1988vdp}. This is another possible connection between particle physics and cosmology.}  can be determined through experiment and observation, however, there is no present fundamental theory that accounts for the many numerical values of the parameters in these matrices.

This is not the only mystery. We know that the oscillation phenomenon indicates that neutrinos have {\index mass eigenstate}s that are combinations of their flavor eigenstates. This further implies that at least two of the neutrinos have non-zero masses. The oscillation periods are determined by parameters involving the differences between the squares of the neutrino masses, e.g., ($m_{2}^{2}$ - $m_{1}^{2}$), so that the individual masses themselves are not determined by the oscillations. 

Although we do not know the masses of each of the three neutrino mass eigenstates, we know that they are much lighter than those of the other known particles. Recently the Karlsruhe Tritium Neutrino experiment ({\index KATRIN}), by studying the endpoint energy of the electron energy spectrum from tritium $\beta$-decay, has placed an upper limit on the neutrino-mass scale of 0.8 eV at a 90\% confidence level~\cite{KATRIN:2019yun}. 

Astronomical observations bearing on cosmological parameters play an important role in
neutrino physics. By comparing data from the {\index Planck satellite} with simulations of the development of structure in the universe one finds an upper limit on the sum of the neutrino mass states of $m_{\nu,tot} < 0.12$ eV at the 95\% confidence level~\cite{Planck:2018vyg}. This indirect result, which is almost an order of magnitude better than the present laboratory result, shows the connection between neutrino physics and cosmology. 

The existence of neutrino masses that are extremely small compared to those of other
particles poses another mystery. Such small, but non-zero, masses are not easily explicable in the context of the {\it {\index standard model}} of particle physics. In that scenario, other fermions acquire masses owing to the strength of their interactions with {\index Higgs boson}s, although the values of the fermion masses are not themselves determined.\footnote{An important consequence of the form of the fermion-Higgs-fermion interaction is its linear dependence of the fermion mass. The larger the mass the stronger this interaction.} However, neutrino masses are too small to comfortably fit into this scheme. Also, a comparison between the numerical values in the {\index PMNS matrix} with those in the {\index CKM matrix} shows that the there is much more neutrino flavor mixing than there is mixing between  quark flavors.  The explanation of these neutrino mass characteristics hint at physics that most probably lies {\it beyond the standard model.} 

Not only do we not know the exact individual masses of the neutrino mass eigenstates themselves, we also do not know whether they follow the same hierarchy as their associated leptons. Furthermore, we do not know the basic nature of the neutrino. We do not know if the neutrino wave functions are described as {\it {\index Majorana}} type (two-component) or {\it {\index Dirac}} type (four-component). A Majorana neutrino is its own antiparticle, whereas a Dirac neutrino has a separate antimatter partner. 
There has also been much interest in searching for the speculative existence of so-called {\it {\index sterile}} neutrinos that may play a role in cosmology and in the explanation of neutrino masses. However, the results of recent sterile neutrino searches with the {\index MicroBooNE} detector been negative.

Since massive neutrinos do not travel at speed of light, in principle, an observer can go faster than the neutrino and look back at it.  The neutrino would then appear as right-handed to the observer.  If this is a {\it distinct Dirac fermion} that is a right-handed neutrino field, it will not participate in any of the standard model left-handed gauge interactions. On the other hand, we know that antineutrinos are right-handed. If this right-handed neutrino is an anti-neutrino, then there is no fundamental distinction between neutrinos and anti-neutrinos. Thus, we are then really talking about a {\index Majorana neutrino}.

The existence of Majorana neutrinos is searched for by looking for {\index neutrinoless double $\beta$-decays}. Such decays violate the conservation of lepton number that holds for other interactions. As of this writing no neutrinoless double beta decays have been found. However, if there are right-handed heavy neutrinos that are of Majorana character, the extremely small mass of the known left-handed neutrinos may be explained by the so-called {\it {\index seesaw mechanism}}. This mechanism explains the small neutrino mass by the exchange of such heavy neutrinos with a Majorana neutrino having a mass of the order of the of {\it {\index Grand Unification}} (GUT) scale, $\cal{O}$$(10^{16})$ GeV, where the strength of the strong, weak and electromagnetic forces become equal. GUT theory might provide another connection with cosmology, namely by hinting at a possible explanation for the {\index baryon asymmetry of the universe} that was left over from the big bang, i.e., the fact that the visible matter in the universe vastly outnumbers antimatter. This shows another connection between neutrino physics, astrophysics, and cosmology.

There are important other fundamental connections between these subjects, as well as astrophysics in general. Because of the astronomically long baselines and very high energies of astrophysical neutrinos, they are ideal probes for testing {\index Lorentz  invariance} and {\index $\cal{CPT}$ ($\cal{CP}$ times time reversal) invariance} and other new physics, i.e., the search for new physics beyond the {\index standard model} of particle physics. The interaction rate for very high energy astrophysical neutrinos interacting with nuclei in a neutrino telescope can inform us about the internal structure of nucleons. All of these topics show that a joint understanding of both neutrino physics and astrophysics is required in order to reach a deeper understanding of cosmology and of fundamental physics itself.

\newpage 
 
\section{Cosmic Neutrinos}

{\index Extragalactic astronomy} at the highest energies observed must be performed by studying neutrinos rather than photons because the universe is opaque to photons at energies above $\sim 10^{15}$ eV. In making observations of neutrinos at energies in excess of $10^{19}$ eV, one can deduce information about the distribution and time history of {\index extragalactic cosmic rays} which were accelerated to energies $\sim 10^{20}$ eV. 

Astrophysical neutrinos of high energy are expected to be produced in astronomical sources such as {\index supernovae}, {\index $\gamma$-ray bursts}, and {\index active galactic nuclei}. Neutrinos are produced in hadronic interactions and {\index photomeson production interactions} followed by charged pion decays. Electron neutrinos are produced by {\index $\beta$-decays} of {\index cosmic-ray nuclei} or secondary neutrons.

{\index Astrophysical neutrinos} that come from {\index extragalactic sources}, unattenuated during propagation, provide the highest energies and the longest baselines for exploring neutrino physics. Among the physics topics thus opened for exploration are high energy {\index neutrino-nucleon cross sections}, possible neutrino decay, possible neutrinos from the decay or annihilation of putative {\index dark matter} particles, and tests of fundamental physics such as {\index Lorentz symmetry} and {\index $\cal{CPT}$ symmetry}. 

In 1987 a burst of neutrinos with energy greater than 7 MeV was detected on Earth, having been produced in the supernova designated as {\index Supernova 1987A}. This object was located in the {\index Large Magellanic Cloud}, a small companion galaxy to the {\index Milky Way} approximately 50 kpc from Earth. The neutrino burst from SN1987A was simultaneously detected at neutrino detectors on three continents, the {\index Kamiokanda II} neutrino detector in Japan, the {\index IMB} (Irvine--Michigan--Brookhaven) neutrino detector in the United States, and the {\index Baksan detector} in Russia. In all, a total of 24 neutrinos from SN1987A were detected. This event marked the beginning of {\index non-solar neutrino astronomy}. 

In April 2013 the {\index IceCube Neutrino Observatory} Collaboration, led by {\index Francis Halzen}, published an observation of two $\sim$PeV energy neutrinos of cosmic origin dubbed {\it Bert} and {\it Ernie} after puppet characters on a popular American children's television program~\cite{IceCube:2013low}. In the following years, dozens more detections of cosmic high energy neutrinos were made by the IceCube collaboration. In 2017 the IceCube Neutrino Observatory detected a 290 TeV neutrino putatively associated with a {\index $\gamma$-ray flare} from the {\index blazar} TXS0506+56 observed by the Fermi Gamma-ray Space Telescope~\cite{IceCube:2018dnn}.  In November of 2022 the IceCube Collaboration reported the results of a ten-year search for neutrinos from astrophysical $\gamma$-ray sources, revealing the detection of 79 neutrinos from the vicinity of the active galactic nucleus (AGN) known as NGC 1068 with a random probablility of $\sim 10^{-5}$~\cite{IceCube:2022der}.

IceCube has also detected an event consistent with the production of a $W^{-}$ {\index {\it Glashow resonance}}~\cite{Glashow:1960zz} initiated by an astrophysically produced $\sim$6.3 PeV $\bar{\nu_e}$ interacting with an electron in the ice in the detector~\cite{IceCube:2021rpz}, i.e., $\bar{\nu_e} + e \rightarrow W^{-} \rightarrow$ {\it mostly {\index hadronic shower}}. Studies of astrophysical electron antineutrinos at the {\index Glashow resonance} energy may provide a test for cosmic antimatter.

The high energy neutrinos of astrophysical origin detected by IceCube have a roughly isotropic large-scale distribution on the sky. The fact that they are isotropically distributed, and not primarily confined to the plane of the Milky Way, indicates that they are overwhelmingly of extragalactic origin. The vast majority of them are presently not associated with known astrophysical sources. The mystery of their origin is another ongoing problem to be solved. However, we do have some clues and guidelines. These neutrinos must be the result of {\index hadronic interactions}, followed primarily by the decay of charged {\index pions}. The pions are made by interactions that must involve high energy nuclei, overwhelmingly protons, in the universe. Thus, neutrino sources must accelerate protons to very high energy. The protons can then interact either with ambient photons or with other nuclei to produce pions. The concurrent production of neutral pions produces high energy $\gamma$-rays. Such $\gamma$-rays can also be produced by {\index proton synchrotron radiation} in the magnetic fields of such sources. 

There is an important difference between the high energy neutrinos and $\gamma$-rays in astrophysical sources. The $\gamma$-rays can either be annihilated or lose energy though electromagnetic interactions. Neutrinos participate only via weak interactions and are therefore immune from electromagnetic interactions. These fundamental differences between the physics of $\gamma$-rays and neutrinos implies that {\it {\index hidden neutrino sources}} that are not readily detectable via photon emission may exist, possibly explaining the existence of neutrinos from sources in regions of the sky for which no {\index $\gamma$-ray sources} have been identified.

In 1965 the 2.7 K cosmic microwave background (CMB) was discovered by {\index Arno Penzias} and {\index Robert Wilson}~\cite{Penzias:1965wn}, for which they received the  Nobel Prize in 1978. Photomeson interactions of ultrahigh energy cosmic rays with photons of the CMB were predicted to attenuate cosmic rays with energies above $\sim$70 EeV by {\index Kenneth Greisen}~\cite{Greisen:1966jv} and independently by {\index Georgiy Zatsepin} and {\index Vadim Kuz'min}~\cite{Zatsepin:1966jv} shortly thereafter. The decay of the resulting charged pions would then be expected to be a source of neutrinos with energies mainly in the EeV energy range. Space-based detectors are now being planned and constructed to search for such ultrahigh energy neutrinos.

\section{Outline of the Following Chapters}

In the following chapters we will discuss many of these topics and questions in detail. We will examine the known properties of neutrinos and their connections with astrophysics and cosmology. We will also discuss the various techniques for neutrino detection, including many present and proposed future neutrino detectors capable of observations of neutrinos produced in astrophysical sources. This includes neutrinos of ultrahigh energy (above $10^{17}$ eV), as well as neutrinos that might be produced by presently unknown physical processes. 

In the next chapter the physical properties of neutrinos and the physical processes involving neutrinos will be addressed.
Chapter 3 treats the role of neutrinos in cosmology and in the search for the dark matter that makes up 27\% of the universe.
Chapters 4 through 7 discuss the various means for detecting astrophysical neutrinos over the energy ranges from MeV energy
to $10^{11}$ GeV energy. Chapters 8 through 11 discuss the theories of the production of neutrinos in astrophysical sources. Chapter 12 explores the roles that neutrinos can play in probing aspects of fundamental physics, viz., Lorentz symmetry and $\cal{CPT}$ symmetry.

\bibliographystyle{ws-rv-van}
\bibliography{Chapter1}

\vspace{1.0cm}
\end{document}